# MARKOV CHAINS APPLIED TO PARRONDO'S PARADOX: THE COIN TOSSING PROBLEM

Xavier Molinero[1] and Camille Mégnien[2]

[1]Department of Mathematics, Universitat Politècnica de Catalunya · BarcelonaTECH, Terrassa, Spain
xavier.molinero@upc.edu
[2]School of Mathematics and Statistics, Barcelona, Spain
cmegnien@me.com

## ABSTRACT

*Parrondo's paradox was introduced by Juan Parrondo in 1996. In game theory, this paradox is described as: A combination of losing strategies becomes a winning strategy. At first glance, this paradox is quite surprising, but we can easily explain it by using simulations and mathematical arguments. Indeed, we first consider some examples with the Parrondo's paradox and, using the software R, we simulate one of them, the coin tossing. Actually, we see that specific combinations of losing games become a winning game. Moreover, even a random combination of these two losing games leads to a winning game. Later, we introduce the major definitions and theorems over Markov chains to study our Parrondo's paradox applied to the coin tossing problem. In particular, we represent our Parrondo's game as a Markov chain and we find its stationary distribution. In that way, we exhibit that our combination of two losing games is truly a winning combination. We also deliberate possible applications of the paradox in some fields such as ecology, biology, finance or reliability theory.*

## KEYWORDS

*Parrondo's paradox, Markov chain, Engineering Decision Making, Maintenance and Evolution*

## 1. INTRODUCTION

Since we start playing, we start asking how to win. Game theory is a quite recent discipline that emerged around the 19th century. Game theory is essentially the study of mathematical models of strategic interaction among rational decision-makers. In other words, Game theory help us to find a winning strategy based on *mathematical thinking*. Game theory does not stop at gambling, there is a wide spectrum of applications in social science, biology or computer science. This is why it is an important field of mathematics. In this paper we will focus on a specific paradox of game theory: the Parrondo's paradox [1]. This paradox states that there exists two losing games such as we can combine those games into a winning game. At first glance, this paradox seems counterintuitive and we can easily see how it made such noise in the game theory field. Everybody jumped on the opportunity and tried to find revolutionary applications of the paradox, and still is. However, could Parrondo's paradox really revolution the way we gamble or invest and so on?
Related work about background of Parrondo's paradox is [31, 32], among others.

We do a deep study of this paradox. First, we define the Parrondo's paradox and see some examples of Parrondo's paradox to have a good first understanding of it. Section 3 simulates one of those examples and try different combinations of the considered games. Later we introduce some definitions and results about Markov chains [2] to help us with the exhaustive study our specific example. Section 5 applies the given concepts of Markov chains to

Parrondo's paradox for coin tossing. We also go over the many possible applications of the paradox. Next, we present our discussion and conclusions in Section 7. Finally, we expose the used materials and methods.

## 2. WHAT IS PARRONDO'S PARADOX: A FEW EXAMPLES

First, we stablish the formal concept of Parrondo's paradox to consider some known examples.

**Definition 1.** The Parrondo's paradox is defined as: There exists two losing games such as specific combinations of these two games lead to a winning game.

Here we introduce some known examples of those games.

**Example 1: A simple coin game [3,4].** In Game A, you simply lose one euro every time you play. In Game B, you count how much money you have left. If it is an even number, you win 3 euros. Otherwise, you lose 5 euros.

Suppose you begin with 100 euros in your pocket. If you start playing Game A exclusively, you will lose all your money in 100 rounds. Similarly, if you decide to play Game B exclusively, you will also lose all your money in 100 rounds.

However, consider playing the games alternatively, starting with Game B, followed by A, then by B, and so on (BABABA...). Then we will win 2 euros every two games!

Thus, even though each game is a losing proposition if played alone, the sequence in which the games are played can lead to a profit.

**Example 2: Saw-tooth [5,6].** Imagine a rack almost vertical, which is going down at constant speed. We will put a ball on it. The ball will be stabilized by the teeth of the rack and thus it will go down with it. We will say that it is a failure if the ball reaches ground and success if it reaches the ceiling. Hence, this rack is a losing game for the ball.

Now suppose we have a second rack. This one is globally going down but alternates between small ascending and descending movements. This game is also losing, as the ball will reach the ground.

However, if we combine those two racks in such a way that the ball will go from one rack to the other at the right time, we could bring the ball to the ceiling. Hence making a winning game from two losing games.

**Example 3: Roulette [7].** Let us consider a game played on a roulette table. The wheel has slots for 0 and the numbers 1 to 36. The zero is coloured green, and half of the numbers from 1 to 36 is coloured black, and the other half in red. All numbers are equally likely to be chosen. We say that when zero turns up the casino always wins.

The first game is the following: You bet on red or black and win one euro if that color turns up and lose one euro otherwise. You will win with probability 18/37<0.5. Hence, this first game is a losing game.

Now the second game is a bit more complicated: If your capital is a multiple of 3 and one of the numbers 1,2 or 3 turns up, you win one euro, otherwise you lose one euro. You will win with probability 3/37. If your capital is not a multiple of 3 and if the outcome is between 1 to 28, you win one euro, otherwise you lose one euro. You will win with probability 28/37. This is a bit more complicated to see but this is also a losing game.

We can combine those two losing games into a winning one, thus exhibiting Parrondo's paradox.

**Example 4: Coin tossing [8].** This example is similar to the previous one, but we consider coin tossing instead of a roulette table. It is also similar to example 1, but a bit more complex.

We have two games, A and B. Game A consists of flipping a coin. We win one euro if it lands on head and loose one euro otherwise. However, this is a biased coin and the probability of landing head is 0.5-$\alpha$, where $\alpha >0$. Hence, the probability of losing is 0.5+$\alpha$). We will consider small values of alpha ($\alpha<0.1$). Clearly, this game is not fair.

Let us look at the second game: Game B is a bit more complex. There is two different coins, let be coin 1 and coin 2. The first coin is really biased and lands head with probability 0.1-$\alpha$ and tails with probability 0.9+$\alpha$. For the second coin, it lands head with probability 0.75-$\alpha$ and tails with probability 0.25+$\alpha$. Since we will consider small $\alpha$ the second coin will be preferable. We will choose which coin to toss with a specific rule: if the current gain of the player is a multiple of M (for M an integer) we will play coin 1, otherwise we play coin 2. Note that we have the same principle as game A: if either coin lands head we win one euro otherwise we lose one euro. This game is also a losing game.

Now, a strange thing happens, if we actually combine those two games, we can get a winning game! Clearly, we can see that if we play game A when the accumulated gain is a multiple of 3 and game B otherwise, we will be winning (since we will often play coin 2). But if we combine randomly those two games, we also get a winning one.

We mostly work on this last example (Example 4: Coin tossing). First, we simulate game A and game B. Then we will consider different combinations of those two games.

## 3. SIMULATION OF EXAMPLE 4: COIN TOSSING

First, we are going to check if game A and B are actually losing game. We are going to simulate those games using the software R (you can find more information on R at [9]), and usinf a 2.3 GHz Intel Core i5 dual core processor. Each chunk of the code has polynomial time and we have a total running time of 37.46 seconds. You can refer to the authors to get the code implementation in detail. For the simulation we chose $\alpha=0.005$ and M=3. We decided to simulate 50,000 plays, as it is clear enough to see the results. Indeed, Figure 7 shows a clear trend with that much plays.

### 3.1. Game A

Remember game A consisted of tossing a coin. If the coin lands on head we win one euro, otherwise we lose one euro. The probability of landing head is 0.495 and the probability of landing tail is 0.505. Figure 1 shows the simulation of the profit for game A for 50,000 plays.

### 3.2. Game B

Remember game B consisted of the following: we have two coins 1 and 2. If the current profit is a multiple of 3, i.e., M=3, we toss coin 1, otherwise we toss coin 2. Now we win one euro if the coin lands head, we lose one euro otherwise. The probability for coin 1 to land head is 0.095. The probability for coin 2 to land head is 0.745. Figure 2 presents the simulation of the profit for game B for 50,000 plays. We can clearly see that both game A and B are losing in the long run.

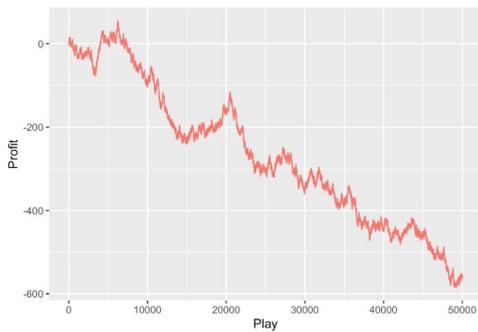
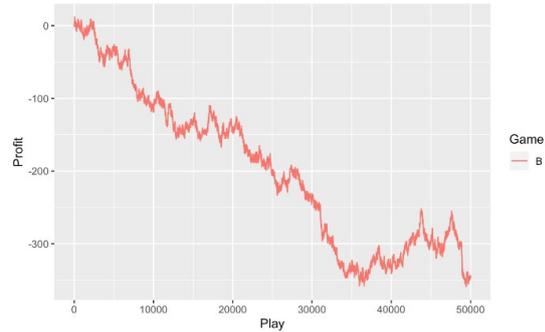

Figure 1. Profit of Game A over 50,000 plays   Figure 2. Profit of Game B over 50,000 plays

Next, we consider some combinations of the games A and B to see if we can get a winning combination.

### 3.3. Game ABABAB

Consider the combination "ABABAB" repeatedly, i.e. we play game A, then game B, then game A and so on. The simulation gives us the results shown in Figure 3. The combination ABABAB is clearly a losing combination in the long run.

### 3.4. Game AABAABAAB

We play the combination "AABAAB" repeatedly. Now a surprising thing happens (see Figure 4): We have a clear profit in the long run if we play the combination "AABAABAAB" repeatedly. Indeed, we can see that with that combination we will often play game B when the current gain is not a multiple of M, and thus we will use the (winning) coin 2.

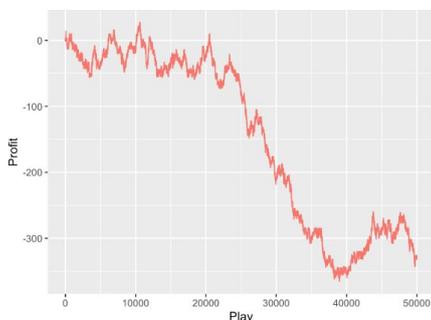
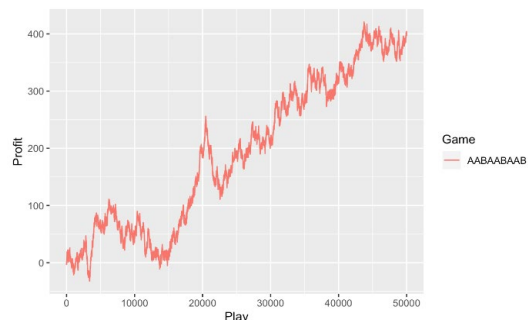

Figure 3. Profit of Game ABABAB over 50,000 plays   Figure 4. Profit of Game AABAAB over 50,000 plays

### 3.5. Game BBBABBBA

The simulation "BBBABBBA" repeatedly give us a losing winning combination (see Figure 5).

### 3.6. Game ABBABB

Figure 6 shows the case of "ABBABBABB" repeatedly. This combination is a winning one.

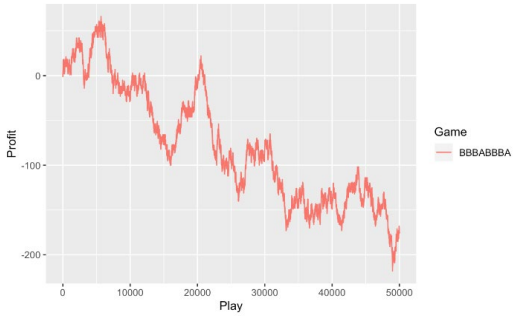
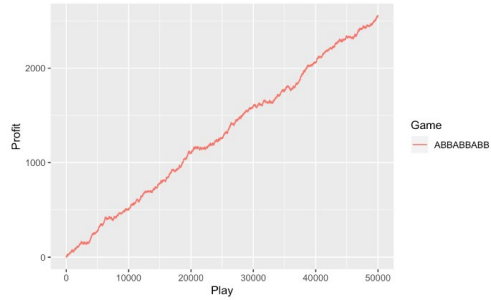

Figure 5. Profit of Game BBBABBBA over 50,000 plays

Figure 6. Profit of Game ABBABB over 50,000 plays

### 3.7. Game random combination

Now, let us try to randomly chose a game at each turn (we chose game A or B with equal probability). We will simulate four random combinations shown in Figure 7.

This result is very surprising; by combining at random those games, we get a profit!
Let us look at the more profitable combination, we will make choices in terms of the actual gain, this is the best possible combination, but it requires choosing a game at each step in terms of our current profit. If our profit is a multiple of M we will choose game A otherwise game B.

### 3.8. Game knowing current gain

Now we are going to choose which game to play in terms of our current profit. We will make the most profitable choice, i.e., at each round we will toss the best coin. Hence, if our profit is a multiple of 3, we will choose to play game A since we will toss the coin with probability of winning of 0.495. If our profit is not a multiple of 3, we will choose to play game B, as we will toss the most profitable coin: coin 2 (probability of winning of 0.745). In that way, we will never toss the coin 1, which is the worst coin (probability of winning of 0.095). Figure 8 shows such profit over 50,000 plays.

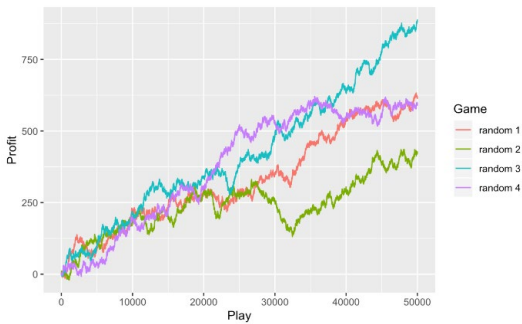
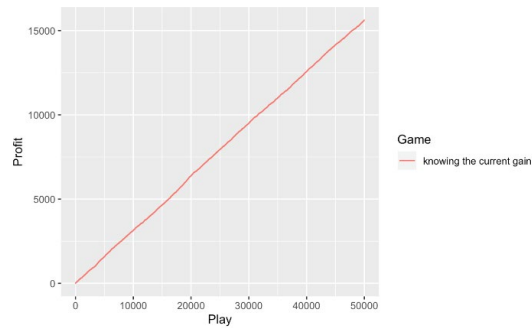

Figure 7. Profit of Game A/B at random over 50,000 plays

Figure 8. Profit of Game "knowing the current gain" over 50,000 plays

### 3.9. Comparing results

Figure 9 plots previous simulations together, and Figure 10 plots all results without the game "knowing the current gain" to see clearly the profits.

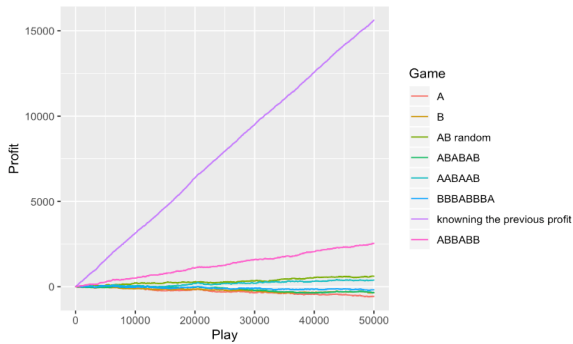
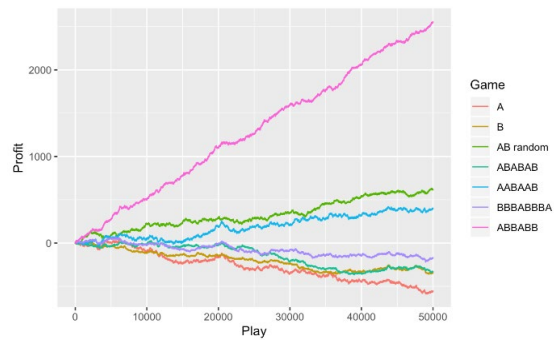

Figure 9. Profit of the all the different game combinations (including "knowing the current gain" over 50,000 plays

Figure 10. Profit of all the different game combinations over 50,000 plays

During this simulation we considered the game where we could choose the game we are going to play in terms of our current gain. In that way, every time we had a profit that is a multiple of M we avoided the game B, since it will lead to the coin 1, which is the worst possible coin in the game. We considered this game as a reference, but we are not going to study it since we prefer games where the strategy is known beforehand and does not change at each step.

Hence, if we look at the last graph, we can see the profit over 50,000 plays of the games A, B, a random alternation of the two games and specific combinations of the two.

First, we see that indeed, game A and B are losing games on their own. We will compute later the expectation of those games.

Then we see that not all combinations of game A and B are winning ones; in fact, the combinations "ABABAB" and "BBBABBBA" are losing combinations, and the combinations "AABAAB" and "ABBABB" are winning. We know that the perfect combination is the one we described above about choosing which game to play in terms of the current gain. Hence, the more the specific combinations are close to this choice, the greater the profit. This is why we get such a high profit for the combination "ABBABB"; it is really close to the one with choices according to profit. Indeed, we can see that we will often play coin 2 of game B.

Now, an interesting thing is that if we choose at random between game A and B, we get a winning game! This result is the most intriguing one and we will work on the math behind it to understand how it can be.

In conclusion, we see that from the two losing games A and B we can get a winning combination. Moreover, a random combination of those two games is also winning.

## 3. MARKOV CHAINS

We have seen in detail one example of the Parrondo's paradox and simulated it. We have seen that this paradox happens, but now we pretend to study how it works. We are going to look at the mathematics behind it to better understand this paradox, for that we will use the theory about Markov Chains. The following is based mostly on those references [2,10–14]. The interested reader can also see some applications to Markov Chains, related with random walks and similar examples that we consider here, in references [15,16]

### 4.1. Introduction to Markov Chains

First, we introduce the Markov chain definition.

**Definition 2**. Markov Chain [14]. Given a stochastic process, i.e. a collection of random variables $X = \{X_t : t \in T\}$, defined on a common probability space taking values in a common set S and indexed by a set T often either N or $[0, +\infty)$ (usually it is thought as time). Then, in order to have a Markov chain, this stochastic process must follow the Markov property:
The conditional probability distribution of future states of the process (conditional on both past and present states) depends only upon the present state, not on the sequence of events that preceded it,
$$\mathbb{P}(X_{n+1} = s_{n+1} | X_n = s_n, X_{n-1} = s_{n-1}, X_{n-2} = s_{n-2}, \ldots) = \mathbb{P}(X_{n+1} = s_{n+1} | X_n = s_n)$$
We define $S = \{s_1, s_2, \ldots, s_n\}$ the set of states. The process starts in one of those states and we move from one state to the other, we call such a move a step.

**Example 5**: A Markov chain. Consider the behaviour of a regular customer of a bookstore. Each day, this customer can do three different actions. Either he does not go into the bookstore (N), he goes to the bookstore but does not buy any books (G) or he goes into the store and buy at least one book (B). Hence, we obtain the following state space: S={N,G,B}. We suppose that the first day the customer has a probability of 0.5 to visit the bookstore without buying, and a probability of 0.5 of buying at least one book.

Imagine we have the following:
- When the customer does not go into the bookstore a day, he has a chance of 25% of still not going the next day, 50% chance of visiting without buying any book and 25% chance of buying a book.
- When the customer visits the bookstore without buying any book, he has a 50% chance of to visit the next day without buying and 50% of buying at least one book.
- When the customer buys at least one book, he has 33% chance of not coming the next day and 33% chance of coming but not buying anything and a 34% chance to buy again.

Example 5 is clearly a Markov chain as the action of the customer the next day only depends of the actions of today and not the day before.
Now let us introduce some definitions that we will need for our study.

**Definition 3 [14]**. The initial probability distribution of the Markov chain is the probability distribution of the chain at time $n = 0$: $\mathbb{P}(X_0 = s) = q(s) \ \forall s \in S$.

**Definition 4. Transition probability kernel [14]**. The probability to move from state $s_{n+1}$ to state $s_n$ is
$$\mathbb{P}(X_{n+1} = s_{n+1} | X_n = s_n) = p_{s_n s_{n+1}} \ \forall (s_{n+1}, s_n) \in S \times S$$
and is called the transition probability kernel.

Assume we have a finite number N of possible states in S: $S = \{s_1, s_2, \ldots, s_N\}$. In order to simplify notations and computations, it is easier to represent our Markov chain with vectors and matrices; we will describe the initial probability distribution by a row vector **q** of size N and the transition probabilities can be described as a square matrix **P** of size $N \times N$:
$$(q)_i = q(s_i) = P(X_0 = s_i)$$
$$(P)_{i,j} = p_{s_i s_j} = \mathbb{P}(X_{n+1} = s_j | X_n = s_i) \ (independent \ of \ n)$$
Hence, we can express each probability distribution at any step as a row vector: $(q_n)_i = q_n(s_i) = \mathbb{P}(X_n = s_i)$

**Example 6.** Coming back to our previous example we have $S = \{N, G, B\}$.
The initial probability distribution is $q = (0, 0.5, 0.5)$.
The transition matrix is $\mathbf{P} = \begin{matrix} N \\ G \\ B \end{matrix} \begin{bmatrix} 0.25 & 0.5 & 0.25 \\ 0 & 0.5 & 0.5 \\ 0.33 & 0.33 & 0.34 \end{bmatrix}$, where the rows represent the action of the present day and the columns the ones of tomorrow.

### 4.2. Computation with Markov chain

Now we are interested in the probability that given the chain is in state $i$ today, it will be in state $j$ two days from now. We denote this probability by $p_{ij}^{(2)}$.

**Example 7.** In the previous example, we see that if the customer did not go to the bookstore today than the event that he will buy a book two days from now is the disjoint union of the events:
  - A= {The customer does not go into the bookstore tomorrow and he buys a book in two days}
  - B= {The customer goes into the store tomorrow and buys a book the day after}
  - C= {The customer buys a book tomorrow and same in two days from now.}

$P(A) = P(\text{"The customer does not go into the bookstore tomorrow"} | \text{"He doesn't go to the store today"}) \cdot P(\text{"The customer buys a book two days from now"} | \text{"he doesn't go into the store tomorow"})$

Thus:
$$P(A) = p_{11} p_{13}$$
$$P(B) = p_{12} p_{23}$$
$$P(C) = p_{13} p_{33}$$

and
$$p_{13}^{(2)} = P(A) + P(B) + P(C) = p_{11} p_{13} + p_{12} p_{23} + p_{13} p_{33}$$

Now, we realize that this equation is expressed as the dot product between the first row and the third column of **P**. In general, if we suppose there is r states we have:
$$p_{ij}^{(2)} = \sum_{k=1}^{r} p_{ik} p_{kj}.$$

**Theorem 1 [12].** Let **P** be the transition matrix of a Markov chain. The ij-th entry of the matrix $\mathbf{P}^n$ gives the probability that the Markov chain, starting in state $s_i$, will be in state $s_j$ after n steps; i.e. $p_{ij}^{(n)} = (\mathbf{P}^n)_{ij}$.

**Proof of Theorem 1.** Remember the matrix multiplication formula:
$$(A * B)_{ij} = \sum_{k=1}^{r} (A)_{ik} * (B)_{kj}.$$
By induction, we have the following.
Initial step: $p_{ij}^{(2)} = \sum_{k=1}^{r} p_{ik} p_{kj} = \sum_{k=1}^{r} (P)_{ik} (P)_{kj} = (P * P)_{ij} = (P^2)_{ij}$, by the previous example, which corresponds to the ij-th entry of the matric $\mathbf{P}^2$. Heredity: Suppose the theorem is true for n, let us show it for n+1.
Remark that in order to get from state $s_i$ to state $s_j$ in $n+1$ steps is the same as going from state $s_i$ to intermediate state $s_k$ in n steps and then going from state $s_k$ to state $s_j$ in one step (this corresponds to $p_{ik}^{(n)} * p_{kj}$), and this for all k. Hence by the law of total probabilities:
$$p_{ij}^{(n+1)} = \sum_{k=1}^{r} p_{ik}^{(n)} * p_{kj}$$
$$= \sum_{k=1}^{r} (P^n)_{ik} (P)_{kj} \text{ (by induction hypothesis)}$$
$$= (P^n * P)_{ij} = (P^{n+1})_{ij}$$

It verifies the heredity and, thus, the theorem is true for any values of n. □

**Theorem 2 [12]**. Let **P** be the transition matrix of a Markov chain, and let **q** be the probability vector, which represents the starting distribution. Then the probability that the chain is in state $s_i$ after n steps is the i-th entry in the vector: $q_n = q * P^n$, i.e., $\mathbb{P}(X_n = s_i) = (qP^n)_i$

**Proof of Theorem 2.** The proof of the theorem is done by induction: Initialization step for n=1:
$\mathbb{P}(X_1 = s_i) = \sum_{k=1}^{r} \mathbb{P}(X_0 = s_k)\mathbb{P}(X_1 = s_i | X_0 = s_k)$, by the law of total probabilities.
Note that $\mathbb{P}(X_0 = s_k) = (q)_k$ and $\mathbb{P}(X_1 = s_i | X_0 = s_k) = (P)_{ki}$ by definition.
Hence, $\mathbb{P}(X_1 = s_i) = \sum_{k=1}^{r} (q)_k (P)_{ki} = (qP)_k$
Therefore, the theorem is true for n=1.
Heredity: Suppose the theorem is true for n and let us show it for n+1:

$$\mathbb{P}(X_{n+1} = s_i) = \sum_{k=1}^{r} \mathbb{P}(X_n = s_k)\mathbb{P}(X_{n+1} = s_i | X_n = s_k)$$

Note again that $\mathbb{P}(X_n = s_k) = (qP^n)_k$ by induction hypothesis. So, $\mathbb{P}(X_{n+1} = s_i | X_n = s_k) = (P)_{ki}$ by Theorem 1. Hence,

$$\mathbb{P}(X_{n+1} = s_i) = \sum_{k=1}^{r} (qP^n)_k (P)_{ki} = (qP^n * P)_i = (qP^{n+1})_i$$

Hence, the theorem is true for any n. □

**Example 8.** Coming back to our example, we can now compute the probability of each state for the second day (n=1) for our customer:

$$q_1 = qP = (0 \quad 0.5 \quad 0.5) \begin{pmatrix} 0.25 & 0.5 & 0.25 \\ 0 & 0.5 & 0.5 \\ 0.33 & 0.33 & 0.34 \end{pmatrix} = (0.165 \quad 0.415 \quad 0.420)$$

We can also use a graph to represent our Markov chain, see Figure 11.

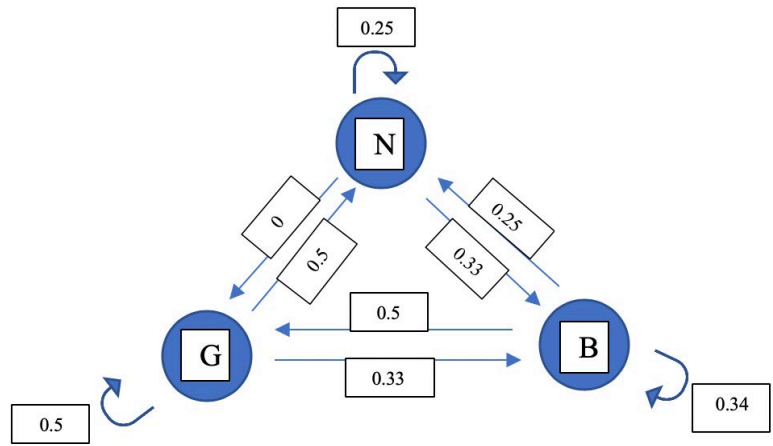

Figure 11. Representation of Example 5 as a graph

Next, we need to introduce additional definitions to stablish our main results.

**Definition 5. Irregular or ergodic Markov chain [12]**. We say that a Markov chain is irreducible or ergodic if it is possible to go from every state to every state (not necessarily in one move).

Observe that if we represent our Markov chain as a graph; the chain is irreducible if the graph is strongly connected.

**Example 9.** Our example about the bookstore (Example 5) is clearly an irreducible Markov chain. We can see it by looking at the graph: it is strongly connected. We also can go from any node to any other node.

**Definition 6. Regular Markov chain [12]**. A Markov chain is call regular if some power of the transition matrix has only positive elements.

Note that in this paper the term "regular" will be used to refer to the preceding definition and never for an invertible matrix.
Intuitively, we can say that a Markov chain is regular if it is possible to go from any state to any state in exactly n steps. We see clearly that every regular chain is irreducible. However, the other way around is not true, see the following example.

**Example 10.** Let P be the transition matrix of a Markov chain:
$$P = \begin{bmatrix} 0 & 1 \\ 1 & 0 \end{bmatrix}$$
Clearly, the chain is irreducible. Figure 12 shows this example as a graph.

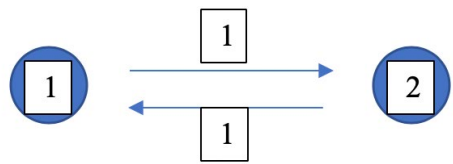

**Figure 12.** Representation of Example 10 as a graph

But the chain is not regular. Suppose that n is odd, then it is not possible to go from state 1 to state 1 in n steps. If n is even, it is not possible either to go from step 1 to step 2 in n steps.

**Example 11.** Is our bookstore example (Example 5) a regular chain? Yes!
Indeed, we can see that it is possible to go from any state to any other state, in two steps. Another way to see it is to compute $P^2$ and see that all entries are positive:
$$P^2 = \begin{bmatrix} 0.145 & 0.4575 & 0.3975 \\ 0.165 & 0.415 & 0.42 \\ 0.1947 & 0.4422 & 0.3631 \end{bmatrix}.$$

**Theorem 3. Fundamental limit theorem for Regular chain [12]**. Let **P** be the transition matrix for a regular chain, then, as n goes to infinity, the powers $P^n$ approach a limiting matrix **W** with all rows the same vector **w**. The vector **w** is a strictly positive probability vector (i.e. the components are all positive and they sum to one).

In order to prove this Theorem, we first need the following lemma:

**Lemma 1.** Let **P** be a transition matrix for dimensions $r \times r$, with no zero entries. Let d be the smallest entry of the matrix. Let **y** be a column vector with r components, let $M_0$ be the largest of those components and $m_0$ the smallest. Let $M_1$ and $m_1$ be the largest and smallest components of the vector **Py**. Then:
$$M_1 - m_1 \leq (1 - 2d)(M_0 - m_0)$$

**Proof of Lemma 1.** First, let us understand what this lemma is saying; if an $r \times r$ transition matrix has no zero entries, and **y** is any column vector with r entries, then the vector **Py** has entries which are "closer together" than the entries are in **y**.

First, note that since each row of **P** is a probability vector, **Py** replaces **y** by averages of its components (with different weights).

The largest weighted average that could be obtained in the present case would occur if all but one of the entries of y have value $M_0$ and the one left entry has value $m_0$, and this one small entry is weighted by the smallest possible weight, namely $d$. In this case, we will obtain the weighted average: $dm_0 + (1-d)M_0$.

The smallest weighted average would be obtained if all the entries of **y** except one have values $m_0$, and the one left has value $M_0$, and $M_0$ is weighted by $d$. We then obtain the average: $dM_0 + (1-d)m_0$

Thus,
$$M_1 \leq dm_0 + (1-d)M_0 \text{ and } m_1 \leq dM_0 + (1-d)m_0.$$
$$M_1 - m_1 \leq dm_0 + (1-d)M_0 - (dM_0 + (1-d)m_0) = (M_0 - m_0)(1-2d).$$
$\square$

Next, we are ready to proof Theorem 3.

**Proof of Theorem 3: The fundamental limit theorem for regular chain.** We will first review the proof of the theorem for the case where **P** has no zero entries. Let **y** be an arbitrary r-column vector, where r is the number of states of the chain. We assume that r > 1. Otherwise it is trivial. Once again, let $M_n$ and $m_n$ be the maximum and minimum components of the vector $P^n y$. The vector $P^n y$ is obtained from the vector $P^{n-1}y$ by multiplying on the left by **P**. As seen in the proof of the lemma each component of $P^n y$ is an average of the components of $P^{n-1}y$. Thus $M_0 \geq M_1 \geq ...$ and $m_0 \leq m_1 \leq ...$

Hence, each sequence is monotone and moreover bounded: $m_0 \leq m_n \leq M_n \leq M_0$. By the monotone theorem, each of these sequences will converge.

Let $M$ be the limit of $M_n$ and $m$ the limit of $m_n$. We know that $m \leq M$. Now, we want to show that $M - m = 0$. This will be the case if $M_n - m_n$ tends to zero.

Let $d$ be the smallest element of **P**. Since all entries of **P** are strictly positive, we have $d > 0$. By our lemma we have:
$$M_n - m_n \leq (1-2d)(M_{n-1} - m_{n-1}) \Leftrightarrow M_n - m_n \leq (1-2d)^2(M_{n-2} - m_{n-2})$$
$$\Leftrightarrow M_n - m_n \leq (1-2d)^n(M_0 - m_0)$$

Since $r \geq 2$, we must have $d \leq 1/2$, so $0 \leq 1 - 2d < 1$. Hence:
$$0 \leq M_n - m_n \leq (1-2d)^n(M_0 - m_0)$$
$$0 \leq \lim_{n \to \infty} M_n - m_n \leq \lim_{n \to \infty} (1-2d)^n(M_0 - m_0)$$

By the squeeze theorem: $\lim_{n \to \infty} M_n - m_n = 0$.

Since any components of $P^n y$ lies between $M_n$ and $m_n$, each component must approach the same number $l = M = m$. This shows that $\lim_{n \to \infty} P^n y = L$. Where $L$ is the column vector such that $L_i = l \; \forall i \in \{1, ..., r\}$.

Now let **y** be the vector with j-th component equal to 1 and all other components equal to zero. Then $P^n y$ is the j-th column of $P^n$. We do this for each j, and we can see that the columns of $P^n$ approach constant column vectors. That is, the rows of $P^n$ approach a common row vector **w**, i.e. $\lim_{n \to \infty} P^n = W$

We are left to show that all entries in **W** are strictly positive. Let **y** be the vector with j-th component equal to 1 and all other to zero. Then **Py** is the j-th column of **P**, and this column has all entries strictly positive (by hypotheses). The minimum component of the vector **Py** was defined to be $m_1$, hence $m_1 > 0$. Since $m_1 \leq m$, we have $m > 0$. Note finally that this value of $m$ is just the j-th component of **w**, so all components of **w** are strictly positive.

The following Theorem give us the probability vector for Markov chains.

**Theorem 4 [12].** Let **P** be a regular transition matrix and $W = \lim_{n \to \infty} P^n$. Let **w** be the common row of **W**, and let **c** be the column vector all of whose components are 1. Then:
  a. **wP=w** and any row vector **v** such that **vP=v** is a constant multiple of **w**.
  b. **Pc=c** and any column vector **x** such that **Px=x** is a multiple of **c**.

**Proof of Theorem 4.** Let be $P^n \to W$. Thus, $P^{n+1} = P^n P \to WP$. But, $P^{n+1} \to W$, hence $WP = W$.
Now, let **v** be such that $vP = v$. Then $vP^2 = vP = v$, and so on: $vP^n = v$. Taking the limit on both sides, we get: $vW = v$.
Let *s* be the sum of the components of **v**. Then $vW = [\sum v_i w_1 \ldots \sum v_i w_n] = \sum v_i [w_1 \ldots w_n] = sw$.
So, $v = sw$.
For the second part, we proceed the same way and obtain that $x = Wx$, by using that fact that all rows of **W** are the same and that $c_i = 1$, we get that **x** is a multiple of **c**. □

**Corollary 1**. There is only one probability vector such that **vP=v**.

Definition 7. Fixed row vector and fixed column vector [12]. *A row vector such that **wP=w** is called a fixed row vector for **P**. Similarly, a column vector **x** such that **Px=x** is called a fixed column vector for **P**.*

**Theorem 5 [12].** Let **P** be the transition matrix for a regular chain and **v** an arbitrary probability vector. Then $\lim_{n \to \infty} vP^n = w$, where **w** is the unique fixed probability vector for **P**.
**Proof of Theorem 5.** By the theorem $\lim_{n \to \infty} P^n = W$, thus $\lim_{n \to \infty} vP^n = vW$.
But, since **v** is a probability vector, its entries sum up to 1, and remember that all rows of **W** are equal to **w**. Hence, we get: $vW = [\sum v_i w_1 \ldots \sum v_i w_n] = \sum v_i w = w$. □

**Theorem 6. Equilibrium Theorem [12].** For an irregular or ergodic Markov chain, there is a unique probability vector **w** such that **wP = w** and **w** is strictly positive. Any row vector such that **vP = v** is a multiple of **w**. Any column vector **x** such that **Px = x** is a constant vector.

**Proof of theorem 6.** This theorem is the same as before but for ergodic chain, not regular.
Let **P** be the transition matrix of an ergodic chain.
Let $P' = (1/2)I + (1/2)P$. This is a regular transition matrix with the same fixed vectors as **P**.
First, let us show that the fixed vectors are the same. Suppose **w** is a fixed vector for **P**, then $wP = w$. Hence, $wP' = (1/2)wI + (1/2)wP = (1/2)w + (1/2)w = w$.
If $wP' = P' \Rightarrow (1/2)w + (1/2)wP = w \Rightarrow (1/2)wP = (1/2)w \Rightarrow wP = w$.
We apply the same argument to the column vector. □

In this section, we have introduced Markov chains. We also showed how to compute with Markov chains, i.e. how to find the next probability distribution. Finally, and most importantly, we found the equilibrium distribution of a regular Markov chain through the fundamental limit theorem for regular chains. In the next part, we will define our Parrondo's games as finite regular Markov chains in order to apply this last theorem and clearly determine how we can make a winning game out of two losing games.

## 5. MATHEMATICAL STUDY OF THE PARADOX

Remember our coin tossing example from Section 2 (Example 4). It consisted of two games A and B. In each game, we had to toss a coin and head corresponded to a win, tail to a loss. $0 < \alpha < 0.1$.

Game A: probability of landing heads=0.5-$\alpha$ tails=0.5+$\alpha$
Game B: Coin 1: probability of getting head= 0.1-$\alpha$, tails=0.9+$\alpha$
    Coin 2: probability of getting heads=0.75-$\alpha$, tails=0.25+$\alpha$
If the current gain is a multiple of M, we play coin 1; otherwise, we play coin 2.
Game AB: Play game A, then B, then A, B, and so on.

We have seen through the simulation (Section 3) that game A and B are losing but game AB is a winning one. Now we want to show this mathematically speaking. We will base our analysis mostly on the articles [1,17,18] and on the theory on Markov chains.

## 5.1. Argument for a general $\alpha$

Now, let us see in general for which $\alpha$ game A and B are losing games and game AB is a winning one. In order to do this analysis, we will consider M=3 (with a bigger or undefined M computations get really complex).

### 5.1.1. Game A

First concerning game A. We can see that game A can consists of independent Bernoulli trials, with probability of success of 0.5-$\alpha$. Clearly, game A will be losing whenever $\alpha > 0$. We can also prove that with Markov chains considering the transition matrix:

$$P_A(\alpha) = \begin{bmatrix} 0 & 0.5-\alpha & 0.5+\alpha \\ 0.5+\alpha & 0 & 0.5-\alpha \\ 0.5-\alpha & 0.5+\alpha & 0 \end{bmatrix}$$

Note that $P_A(\alpha)$ is regular:

$$P_A^2(\alpha) = \begin{bmatrix} 2(0.5-\alpha)(0.5+\alpha) & (0.5+\alpha)^2 & (0.5-\alpha)^2 \\ (0.5-\alpha)^2 & 2(0.5-\alpha)(0.5+\alpha) & (0.5+\alpha)^2 \\ (0.5+\alpha)^2 & (0.5-\alpha)^2 & 2(0.5-\alpha)(0.5+\alpha) \end{bmatrix}$$

Since $\alpha < 0.1$ the entries of $P_A^2(\alpha)$ are strictly positive.
We can apply theorems 3 and 5: $vP_A(\alpha) = v$

$$[v_1 \ v_2 \ v_3] \begin{bmatrix} 0 & 0.5-\alpha & 0.5+\alpha \\ 0.5+\alpha & 0 & 0.5-\alpha \\ 0.5-\alpha & 0.5+\alpha & 0 \end{bmatrix} = [v_1 \ v_2 \ v_3]$$

*knowing that* $v_1 + v_2 + v_3 = 1$. We obtain the following system of equations:

$$(0.5+\alpha)v_2 + (0.5-\alpha)v_3 = v_1$$
$$(0.5-\alpha)v_1 + (0.5+\alpha)v_3 = v_2$$
$$(0.5+\alpha)v_1 + (0.5-\alpha)v_2 = v_3$$
$$v_1 + v_2 + v_3 = 1$$

Which is equivalent, as before, to:

$$v_1 = 1/3$$
$$v_2 = 1/3$$
$$v_3 = 1/3$$

Then, the probability of winning one play in the long term is (by the total law of probabilities):

$P(win|we\ are\ in\ s_1)P(to\ be\ in\ s_1) + P(win|we\ are\ in\ s_2)P(to\ be\ in\ s_2) + P(win|we\ are\ in\ s_3)P(to\ be\ in\ s_3) = (0.5-\alpha) \cdot 1/3 + (0.5-\alpha) \cdot 1/3 + (0.5-\alpha) \cdot 1/3 = \mathbf{0.5 - \alpha}$

Hence, game A is losing whenever this probability is strictly lower than 0.5 i.e. when $\alpha > 0$.

### 5.1.2. Game B

Secondly, we consider game B; for which $\alpha$ is game B a losing game?
Consider the following transition matrix:

$$P_B(\alpha) = \begin{bmatrix} 0 & 0.1-\alpha & 0.9+\alpha \\ 0.25+\alpha & 0 & 0.75-\alpha \\ 0.75-\alpha & 0.25+\alpha & 0 \end{bmatrix}$$

Note that $P_B(\alpha)$ is regular:

$$P_B^2(\alpha)$$
$$= \begin{bmatrix} (0.1-\alpha)(0.25+\alpha)+(0.75-\alpha)(0.9+\alpha) & (0.9+\alpha)(0.25+\alpha) & (0.1-\alpha)(0.75-\alpha) \\ (0.75-\alpha)^2 & (0.1-\alpha)(0.25+\alpha)+(0.75-\alpha)(0.25+\alpha) & (0.9+\alpha)(0.25+\alpha) \\ (0.25+\alpha)^2 & (0.75-\alpha)(0.1-\alpha) & (0.75-\alpha)(0.9+\alpha)+(0.75-\alpha)(0.25+\alpha) \end{bmatrix}$$

Since $\alpha < 0.1$ the entries of $P_B^2(\alpha)$ are strictly positive.
We can apply the theorems 3 and 5.
$$vP_B(\alpha) = v$$
$$[v_1 \quad v_2 \quad v_3] \begin{bmatrix} 0 & 0.1-\alpha & 0.9+\alpha \\ 0.25+\alpha & 0 & 0.75-\alpha \\ 0.75-\alpha & 0.25+\alpha & 0 \end{bmatrix} = [v_1 \quad v_2 \quad v_3]$$

And $v_1 + v_2 + v_3 = 1$.
We obtain the following system of equations:
$$(0.25+\alpha)v_2 + (0.75-\alpha)v_3 = v_1$$
$$(0.1-\alpha)v_1 + (0.25+\alpha)v_3 = v_2$$
$$(0.9+\alpha)v_1 + (0.75-\alpha)v_2 = v_3$$
$$v_1 + v_2 + v_3 = 1$$

Which is equivalent to:
$$v_1 = \frac{5(16\alpha^2 - 8\alpha + 13)}{240\alpha^2 - 16\alpha + 169}$$
$$v_2 = \frac{2(40\alpha^2 + 6\alpha + 13)}{240\alpha^2 - 16\alpha + 169}$$
$$v_3 = \frac{2(40\alpha^2 + 6\alpha + 39)}{240\alpha^2 - 16\alpha + 169}$$

Then, the probability of winning one play in the long term is (by the total law of probabilities):
$P(win|we\ are\ in\ s_1)P(to\ be\ in\ s_1) + P(win|we\ are\ in\ s_2)P(to\ be\ in\ s_2) + P(win|we\ are\ in\ s_3)P(to\ be\ in\ s_3) = (0.1-\alpha) \cdot v_1 + (0.75-\alpha) \cdot v_2 + (0.75-\alpha) \cdot v_3 = (0.1-\alpha) \cdot \frac{5(16\alpha^2-8\alpha+13)}{240\alpha^2-16\alpha+169} + (0.75-\alpha) \cdot \frac{2(40\alpha^2+6\alpha+13)}{240\alpha^2-16\alpha+169} + (0.75-\alpha) \cdot \frac{2(40\alpha^2+6\alpha+39)}{240\alpha^2-16\alpha+169}$

Game B is a losing game whenever this probability is strictly lower than $0.5$ i.e. when
$$\frac{-240\alpha^3 + 144\alpha^2 - 155\alpha + 84.5}{240\alpha^2 - 16\alpha + 169} < 0.5 \Leftrightarrow -240\alpha^3 + 24\alpha^2 - 147\alpha < 0 \Leftrightarrow \alpha > 0$$

Hence, game B is a losing game whenever $\alpha > 0$ (same that game A).

### 5.1.3. Game AB

Finally, for the game AB which $\alpha$ makes it a winning game?
We consider the following transition matrix:
$$Q(\alpha) = \begin{bmatrix} 0 & 0.3-\alpha & 0.7+\alpha \\ 0.375+\alpha & 0 & 0.625-\alpha \\ 0.625-\alpha & 0.375+\alpha & 0 \end{bmatrix}$$

Note that $Q(\alpha)$ is regular:

$$Q^2(\alpha)$$
$$= \begin{bmatrix} (0.3-\alpha)(0.375+\alpha)+(0.7+\alpha)(0.625-\alpha) & (0.7+\alpha)(0.375+\alpha) & (0.3-\alpha)(0.625-\alpha) \\ (0.625-\alpha)^2 & (0.375+\alpha)(0.3-\alpha)+(0.625-\alpha)(0.375+\alpha) & (0.375+\alpha)(0.7+\alpha) \\ (0.375+\alpha)^2 & (0.625-\alpha)(0.3-\alpha) & (0.625-\alpha)(0.7+\alpha)+(0.375+\alpha)(0.625-\alpha) \end{bmatrix}$$

Since $\alpha < 0.1$ the entries of $Q(\alpha)$ are strictly positive.
We can apply the theorems 3 and 5.
$$vQ(\alpha) = v$$

$$[v_1 \quad v_2 \quad v_3] \begin{bmatrix} 0 & 0.3-\alpha & 0.7+\alpha \\ 0.375+\alpha & 0 & 0.625-\alpha \\ 0.625-\alpha & 0.375+\alpha & 0 \end{bmatrix} = [v_1 \quad v_2 \quad v_3]$$

*knowing that* $v_1 + v_2 + v_3 = 1$. We obtain the following system of equations:

$$(0.375 + \alpha)v_2 + (0.620 - \alpha)v_3 = v_1$$
$$(0.3 - \alpha)v_1 + (0.375 + \alpha)v_3 = v_2$$
$$(0.7 + \alpha)v_1 + (0.625 - \alpha)v_2 = v_3$$
$$v_1 + v_2 + v_3 = 1$$

Which is equivalent to:

$$v_1 = \frac{320\alpha^2 - 80\alpha + 245}{960\alpha^2 - 32\alpha + 709}$$
$$v_2 = \frac{320\alpha^2 + 24\alpha + 180}{960\alpha^2 - 32\alpha + 709}$$
$$v_3 = \frac{320\alpha^2 + 24\alpha + 284}{960\alpha^2 - 32\alpha + 709}$$

Then, the probability of winning one play in the long term is (by the total law of probabilities):

$$\tfrac{1}{2}[P(\text{win A}|\text{we are in } s_1)P(\text{to be in } s_1) + P(\text{win A}|\text{we are in } s_2)P(\text{to be in } s_2) + P(\text{win A}|\text{we are in } s_3)P(\text{to be in } s_3)] + \tfrac{1}{2}[P(\text{win B}|\text{we are in } s_1)P(\text{to be in } s_1) + P(\text{win B}|\text{we are in } s_2)P(\text{to be in } s_2) + P(\text{win B}|\text{we are in } s_3)P(\text{to be in } s_3)] =$$

$$\tfrac{1}{2}[(0.5-\alpha)\cdot v_1 + (0.5-\alpha)\cdot v_2 + (0.5-\alpha)\cdot v_3] + \tfrac{1}{2}[(0.1-\alpha)\cdot v_1 + (0.75-\alpha)\cdot v_2 + (0.75-\alpha)\cdot v_3]$$

$$= \frac{0.5-\alpha}{2} + \tfrac{1}{2}[(0.1-\alpha)\cdot v_1 + (0.75-\alpha)\cdot(v_2+v_3)]$$

$$= \frac{0.5-\alpha}{2} + \tfrac{1}{2}\left[(0.1-\alpha)\cdot\frac{320\alpha^2-80\alpha+245}{960\alpha^2-32\alpha+709} + (0.75-\alpha)\cdot\left(\frac{320\alpha^2+24\alpha+180}{960\alpha^2-32\alpha+709} + \frac{320\alpha^2+24\alpha+284}{960\alpha^2-32\alpha+709}\right)\right]$$

Game AB is a winning game whenever this probability is strictly greater than 0.5, i.e., when

$$\frac{-1920\alpha^3 + 1056\alpha^2 - 1406\alpha + 727}{1920\alpha^2 - 64\alpha + 1418} > 0.5 \Leftrightarrow \alpha < 0.013109$$

Hence game AB is a winning game for $\alpha < 0.013109$.

## 6. APPLICATIONS OF THE PARADOX

We proved that we could actually combine two losing games into a winning one. So now, what do we do of it? Could we just rush into the closest casino and become rich? The answer is basically no. Indeed, the paradox cannot turn any losing games into winning ones, it follows a specific set of rules. In a Parrondo's game the rules are capital dependent, the game we play depend on the current capital. Game A and B were linked through the gain, playing one game will modify this gain and will have an impact on the next playing game. In casinos, all the games are capital independent; playing one game will not affect another game. In order to apply the paradox in a casino we will need to find three games to represent our three coins. Easily we could find games to model the two losing coins (coin from game A and coin 1 from game B), but we will not find a game in a casino that is winning such as coin 2 from game B.

### 6.1. Application in Physics [19]

Professor Parrondo theorized the paradox is a physicist. He was working on flashing Brownian Ratchets. This is a rather complex device: It consist of a ratchet that is freely to move in one direction only. The ratchet is connected to an immersed wheel in a fluid of molecules at a certain temperature. The molecules will undergo Brownian motion and constitute a heat bath. The impulse from a molecular collision will turn the immersed wheel. Remember that the ratchet can only turn in one direction. Hence the net effect of all the collisions would seem to make the ratchet turned continuously in that direction (see Figure 13).

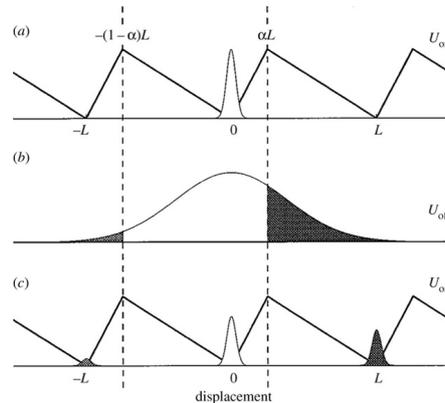

Figure 13. Flashing Brownian Ratchets [5]

We can see it more clearly in this representation of the potentials: (a) represents the potential on (game B), (b) the potential off (game A) and (c) the potential on (game B).

If the potential is switched on (analogous to game B), the Brownian particles will fall into one of the "valleys", as we see in the graph (a), we suppose they fall around 0. Then, if we turn off the potential, the particle will spread out. This is analogous to game A. Finally, if we switch on again the potential, the particles that have been spread out will fall again into the "valleys". But now, they will fall into different valleys. As we can see in the graph (c), some of them stay in the valley around 0, a small amount of them went into the valley around -L, and another part of the Brownian particle went into the +L valley. Hence, we have a net positive displacement of those particles.

To make it simple; the flashing Brownian ratchet is a process that alternates between two states; a one-dimensional Brownian motion and a Brownian ratchet. This system will produce a directed motion. Those two states alone do not result in a directed motion but the combination of the two is. It is a rather complicated process, and Dr. Parrondo introduced Parrondo's games to illustrate it in simpler way. Note that all the other areas of application (others than gambling) are areas of active research and the implication of Parrondo's paradox is here suggested and left to prove.

## 6.2. Application in Ecology [20]

We consider a population that can express two kind of behavior: nomad or colonist. A nomad behavior corresponds to an independent lifestyle; thus, they are not affected by competition or cooperation. Hence, under poor environmental conditions they will go extinct. In the contrary, colonists live together in close proximity and, therefore, they are affected by competition and cooperation, but then they may drain all the resources of their environment and also go extinct. Hence, we can qualify those two behaviors as losing strategies in the "survival game".

The nomad's behavior will be "equivalent" to our previous game A. Concerning the colonist behavior, we have the following: let A be the critical lower capacity and K the carrying capacity. If the actual size of the population is between A and K, the population will grow, and it will decrease otherwise. Moreover, the carrying capacity K changes depending on the populations size (this represents the fact that environmental resources are destroyed) hence making colonist behavior a losing strategy. Thus, the colonist behavior will be our game B. Indeed, the current gain is replaced by the current population size and, depending on its value, we alternate between a winning (if A<population size<K, the population increases) and losing game (otherwise the population decreases).

To recap, game A represents the nomad behavior (a losing strategy), game B represents the colonist strategy if the population size is between A and K the population will increase, otherwise it decreases, M is represented by A and K, the current gain is represented by the population size.

We suppose that the organism can know the amount of environmental resources to them and hence deduce the carrying capacity. In function of the values of this carrying capacity they will change strategies. We will call strategic alternation when the organism changes behavior from colonist to nomad if the carrying capacity is low and from nomad to colonist if the carrying capacity is high.

Simulations have been run based on those two survival strategies, nomad and colonist, the following has been observed: In case of no-behavioral switching, either the two populations go extinct or the nomad goes extinct whereas the colonial survives. In case of behavioral switching, either they both populations go extinct, or they survive through periodic behavioral alternation, or there is a long-term growth through strategic alternation. Most importantly, one aspect of those simulations illustrates the Parrondo's paradox: there were situations where both populations will die if we did not permit behavior-changes but will survive trough **periodic** behavioral alternation.

We exhibited Parrondo's paradox in this ecological situation. This paradox could have also wider application on the field of ecology such as explaining why a destructive specie like the Homo Sapiens can thrive and grow even with limited environmental resources or help us better understand the evolution or extinction of species in general and maybe the emergence of life.

### 6.3. Application in Finance [21–23]

Finance is one of the most active field of research related to Parrondo's paradox as it could lead to the most profit; everyone wants to find out how we can combine two losing investment into a winning one.

For now, we did not find a direct application of the paradox. The reason for that is that Parrondo's paradox is following specific rules for the determination of Game A and B. To find strategies in asset management that will also follow those rules is nearly impossible. However, let us suppose that we are actually able to find an investment bank willing to sell us a security where we could apply the paradox. Let us then assume that we will be able to build a portfolio which will increase its total value while the values of the stocks will decrease individually. In that case we will make money out of worthless stocks, they must be a loophole somewhere. Moreover, note that it is as hard to find stocks that will decrease as stocks that will increase.

Hence, the Parrondo's paradox have not yet find a utility in stock investment but the research is still active. One article [24] states that "rebalanced portfolio diversification can turn individually, money-losing assets into a winning portfolio". Indeed, it shows that by taking separate investments (that were more likely to be losing ones), if we rebalance them to make an equally weighted portfolio, we were more likely to make profits. Hence, exhibiting Parrondo's paradox once more.

Moreover, we have found a correlation between the paradox and "volatility pumping" [22] similar to the rebalanced portfolio diversification. Volatility pumping consists of the following: We have two stocks A and B. Stock A is stable but not so good in the long run and stock B increases but is volatile. The method is to sell both stocks each day. Let N be the total amount of the sale. Then, we buy N/2 of stock A and N/2 of stock B. We repeat this procedure each day. This method will produce a positive profit. However, this is a toy model and it is hard to apply it to the real stock market. Indeed, it is not possible to buy and sell everyday as they will be high transaction costs.

Those preceding applications on finance are "toy models"; we are still waiting on a breakthrough concerning a direct application of Parrondo's paradox in finance.

### 6.4. Application in Reliability theory [25]

Reliability theory is defined as the probability that a device will perform its intended function under specific times and conditions. To exhibit Parrondo's paradox we studied two systems in series, where the first system was less reliable than the second. Then, we modified the first system by randomly choosing its components, i.e. we chose each unit randomly in the same set of components as before. Hence the distributions of the new units were a mixture of the previous units' distributions. Surprisingly, the new system was more reliable than the second one in specific conditions. Therefore, by randomly choosing units from the first, less reliable, system, we obtained a better system than the second one, thus exhibiting the paradox.

### 6.5. Application 5: Biology

Parrondo's paradox has also some possible application in biology. The first one concerns random phase variation [26]. Phase variation is a method for dealing with changing environment, it involves the variation of protein expression. Random phase variation is a phase variation that happen at random time. But phase variation is a losing strategy as it will often results in a maladaptive phenotype. However, it has been observed that random phase variation has better results in the survival of the organism in certain conditions, and it could even be necessary.

We have also studied some applications in oncology [27]. Indeed, the growth of tumors can admit a chaotic behavior in certain regimes. It has been shown that chaotic tumor growth trajectories can be made nonchaotic by modifying some control parameters as drugs. Hence, we could combine chaotic behaviors into a nonchaotic one, thus reflecting the paradox.

Another application in biology concerns the sensors [28]. The biological sensors of an organism permit him to analyze its environment and make decision upon it. We expect organisms with more accurate sensors to be more adaptive to their environment and be better at survival. But actually, under certain conditions, organisms with less accurate sensors tend to be the one surviving. This also exhibits Parrondo's paradox:

We can consider game A as doing nothing, we will say that this is a losing strategy as the environment is changing. Game B will consist of using the bad sensors to make a decision of whether or not to migrate. In Parrondo's game B, we had two coins, one better than the other, those two coins will be represented by the stochastic switching of the environment: in certain condition in the environment less accurate sensors will make better decisions, on other conditions it will do the opposite. Note that game B in general is a losing strategy as it consists of using bad sensors to make decision. However, it has been seen that less accurate sensors tend to survive over more accurate ones and thus expressing Parrondo's paradox.

Finally, we point out that recent studies over COVID-19 use strategies based on Parrado's paradox, see [29, 30].

### 7. DISCUSSION AND CONCLUSIONS

Through this paper we did a complete study of Parrondo's paradox. We first introduced Parrondo's game through different examples all related to gambling. Then, we decided to simulate the coin tossing Parrondo's game. From that simulation we clearly saw that it was possible to combine two losing games into a winning one. Even more surprising, we saw that a random combination of the two losing games leaded to a winning game. After having illustrated the paradox, we did a full mathematical study of it. In order to do that, we first introduced some definitions and theorems related to Markov chains. Notably we proved a specific version of the fundamental limit theorem for regular Markov chain. This theorem allowed us to represent our example as a regular Markov chain and compute its equilibrium distribution. With all that knowledge, we were able to analyze our Parrondo's game and prove that indeed those two losing games can be combined into a winning game. Finally, we saw that, even though in this work we mostly use the examples of gambling games, Parrondo's paradox could also be applied in many other fields such as physics, ecology, finance, reliability theory or biology.

Therefore, although Parrondo's paradox is a "certain" revolution in game theory, it can actually be "easily" explained by using Markov chains.

## ACKNOWLEDGEMENTS

Xavier Molinero has been partially supported by funds from the Ministry of Science and Innovation grant PID2019-104987GB-I00 (JUVOCO) and the Catalan government [2021 SGR 01419 ALBCOM].

Camille Mégnien has been partially funded by the scholarship Swiss-European Mobility Programme.